\documentclass[journal=acscii]{achemso}
\usepackage[T1]{fontenc} 
\usepackage[version=3]{mhchem}
\usepackage{amsmath}
\usepackage{amssymb}
\usepackage{graphicx}
\usepackage{subcaption}
\usepackage{color}
\usepackage{dcolumn}
\usepackage{bm}
\usepackage{miller}
\usepackage{pdfpages}

\newcommand*{\rtten}[1]{\mathbf{\boldsymbol{#1}}}
\newcommand*{\rtvec}[1]{\mathbf{#1}}

\graphicspath{ {Figures/} }
\author{Zeeshan Ahmad}
\affiliation{Department of Mechanical Engineering, Carnegie Mellon University, Pittsburgh, Pennsylvania 15213, USA}
\author{Tian Xie}
\affiliation{Department of Materials Science and Engineering, Massachusetts Institute of Technology, Cambridge, Massachusetts 02139, USA}
\author{Chinmay Maheshwari}
\affiliation{Department of Mechanical Engineering, Carnegie Mellon University, Pittsburgh, Pennsylvania 15213, USA}
\author{Jeffrey C. Grossman}
\affiliation{Department of Materials Science and Engineering, Massachusetts Institute of Technology, Cambridge, Massachusetts 02139, USA}
\author{Venkatasubramanian Viswanathan}
\affiliation{Department of Mechanical Engineering, Carnegie Mellon University, Pittsburgh, Pennsylvania 15213, USA}
\alsoaffiliation[phys]{Department of Physics, Carnegie Mellon University, Pittsburgh, Pennsylvania 15213, USA}
\email{venkvis@cmu.edu}

\title[An \textsf{achemso} demo]
  {Machine Learning Enabled Computational Screening of Inorganic Solid Electrolytes for Dendrite Suppression with Li Metal Anode}

\begin{document}

\begin{tocentry}
\includegraphics[width=8.4cm]{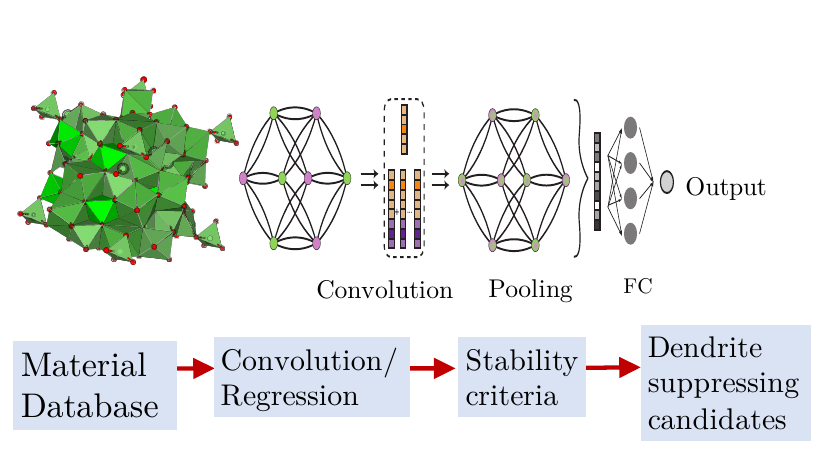}

\end{tocentry}

\begin{abstract}
Next generation batteries based on lithium (Li) metal anodes have been plagued by the dendritic electrodeposition of Li metal on the anode during cycling, resulting in short circuit and capacity loss. Suppression of dendritic growth through the use of solid electrolytes has emerged as one of the most promising strategies for enabling the use of Li metal anodes. We perform a computational screening of over 12,000 inorganic solids based on their ability to suppress dendrite initiation in contact with Li metal anode. Properties for mechanically isotropic and anisotropic interfaces that can be used in stability criteria for determining the propensity of dendrite initiation are usually obtained from computationally expensive first-principles methods. In order to obtain a large dataset for screening, we use machine learning models to predict the mechanical properties of several new solid electrolytes. The machine learning models are trained on purely structural features of the material, which do not require any first-principles calculations. We train a convolutional neural network on the shear and bulk moduli because of the availability of a large training dataset that has low noise due to low uncertainty in their first-principles calculated values. We use AdaBoost, Lasso and Bayesian ridge regression to train the elastic constants, where the choice of the model depended on the size of the training data and the noise that it can handle. Our models give us direct interpretability by revealing the dominant structural features affecting the elastic constants. The stiffness is found to increase with a decrease in volume per atom, increase in minimum anion-anion separation, and increase in sublattice (all but Li) packing fraction.  
Cross-validation/test performance suggests our models generalize well. We predict over twenty mechanically anisotropic interfaces between Li metal and six solid electrolytes which can be used to suppress dendrite growth. Our screened candidates are generally soft and highly anisotropic, and present opportunities for simultaneously obtaining dendrite suppression and high ionic conductivity in solid electrolytes.
\end{abstract}


\section{Introduction}
Increased energy densities of Li-ion batteries are crucial for progress towards complete electrification of transportation \cite{sripad2017lcv, sripad2017metrics,moore2014aviation}.  Among the many possible routes, Li metal anodes have emerged as one of the most likely near-term commercialization options.\cite{xin-bing2017lithium} Coupled with a conventional intercalation cathode, batteries utilizing Li metal anodes could achieve specific energy of $>400$ Wh/kg, much higher than the current state of the art $\sim$250 Wh/kg \cite{cui2017reviving,christensen2011-Liair}. 
Unstable and dendritic electrodeposition on Li metal anode coupled with capacity fade due to consumption of electrolyte has been one of the major hurdles in its commercialization \cite{XuLi2014,TikekarNatEnergy2016,cui2017reviving,aurbach2000,aurbach2002short,steiger2014dendrite, albertus2018status}. For large scale adoption, a stable, smooth and dendrite-free electrodeposition on Li metal is crucial.

Numerous approaches are being actively pursued for suppressing dendrite growth through the design of novel additives in liquid electrolytes \cite{aurbach1996comparative, hirai1994effect,ding2013electrostatic, qian2015high, suo2013solvent,lu2014stable,zhang2017FEC}, surface nanostructuring \cite{Wang17,zhang2014nanorod}, modified charging protocols\cite{mayers2012pulsed,aryanfar2014pulsed}, artificial solid electrolyte interphase or protective coatings \cite{liu2015artificial,yan2014ultrathin,liu2017artificial}, polymers \cite{khurana2014suppression,balsara2012-modadh,yue2016polymer} or inorganic solid electrolytes \cite{janek2016solid, li2014lipon,suzuki2015garnet,manthiram2017solid}. Among these, dramatic improvements in the ionic conductivity of solid electrolytes \cite{Kamaya2011,Kato2016-ssb} have made them extremely attractive candidates for enabling Li metal anodes.

A comprehensive and precise criterion for dendrite suppression is still elusive.  Monroe and Newman performed a dendrite initiation analysis and showed that solid polymer electrolytes with shear modulus roughly twice that of Li could achieve stable electrodeposition \cite{Monroe2005Impact}. In an earlier work, we extended this idea and showed that the criteria for the suppression of dendrite growth gets reversed for inorganic crystalline materials due to the difference in molar volume of \ce{Li+}.  A softer solid electrolyte is required for stability in this case\cite{ahmad2017stability}. It is worth highlighting that this requirement applies only for  dendrite initiation regime and other suppression approaches may be possible for the propagation regime. However, once initiated, dendrite growth is extremely hard to mitigate as pointed out by several studies\cite{diggle1969zinc,monroe2003-dendrite,Monroe2004Effect}. Therefore, it is best to prevent dendrites from initiating to ensure smooth electrodeposition throughout cycling of the battery.

In recent years, high-throughput computational materials design has emerged as a major driver of discovery of novel materials for various applications \cite{curtarolo2013highthroughput,saal2013oqmd}. It typically involves a combination of first-principles quantum-mechanical approaches and database construction and mining techniques. Combined with machine learning methods that bypass the use of expensive quantum mechanical calculations through the use of structural descriptors \cite{pilania2013accelerating, Liu2017materials,gomez2018automatic,deJong2016statistical,isayev2017universal}, one can accelerate the high-throughput screening by several orders of magnitude \cite{fujimura2013method, sendek2017holistic,evans2017predict}. Previous high-throughput screening studies of solid electrolytes have used ionic conductivity, stability and electronic conductivity as screening criteria\cite{sendek2017holistic,fujimura2013method}. However, dendrite suppression capability of solid electrolytes is an additional requirement that needs to be assessed.

Here, we carry out a large-scale data-driven search for solid electrolytes that might be promising candidates for suppressing dendrite growth during the initiation phase with a Li metal anode. We use machine learning techniques to train and predict the mechanical properties of inorganic solids which play a major role in stabilizing the interface. These properties are fed into the theoretical framework which uses the stability parameter \cite{ahmad2017stability,ahmad2017-anisotropy} to quantify the dendrite initiation with Li metal anode. At a mechanically isotropic interface, the screening results predict the crucial role of surface tension in stabilizing the interface since most solid electrolytes are not intrinsically stabilized by the stresses generated at the interface. Hence, surface nanostructuring may be essential to prevent initiation of dendrites for isotropic interfaces. We rank the materials based on the amount of nanostructuring (surface roughness wavenumber) required for achieving a stable electrodeposition. We then performed a stability analysis of over 15,000 anisotropic interfaces between the Li metal and solid electrolyte using the Stroh formalism. This is essential to account for the highly anisotropic mechanical properties of Li\cite{xuLi2017} and texturing of electrodeposited Li at the interface\cite{Cui2017-texturing}. A full anisotropic treatment of the interface reveals over twenty candidate interfaces that are predicted to suppress dendrite initiation. The materials obtained through screening are generally soft and with highly anisotropic mechanical properties. Since softer materials are generally faster ion conductors than stiffer materials due to availability of more volume per atom \cite{wang2015design}, the screened candidates present an opportunity to obtain both desirable mechanical properties and fast ion conduction.



\section{Results and Discussion}
We describe the procedure for screening solid electrolytes and discuss the material candidates obtained based on isotropic and anisotropic criteria separately. First, we briefly review the criteria for dendrite suppression at a metal-solid electrolyte interface to determine the desirable properties of solid electrolytes. We then discuss our machine learning models used for predicting these properties. Finally, we perform material screening based on these properties and discuss the implications.

\subsection{Isotropic Material Screening} 
In solid electrolytes, the mechanical properties at the interface provide an additional degree of freedom for tuning the stability of electrodeposition. Previously, we developed a generalized stability diagram for assessing the stability of electrodeposition at a metal-solid electrolyte interface for isotropic mechanical response \cite{ahmad2017stability}. In these studies, we used the stability parameter first proposed by Monroe and Newman \cite{Monroe2004Effect} to characterize the growth/decay of dendrites with time. The sign of the stability parameter, denoted hereafter as $\chi$, determines whether the electrodeposition is stable or unstable. A positive  $\chi$ implies higher current density at the peaks and lower current density at the valley leading to growth of dendrites while a negative $\chi$ leads to stabilization or suppression of dendrites. The stability parameter is related to the change in the electrochemical potential of the electron $\Delta \mu_{e^-}$ at a deformed interface $z=f(x)$ between the metal anode and the electrolyte (Fig. S1). It is convenient to compute properties of the interface in Fourier space with $f(x)=\int \mathrm{d}k [f_1(k) \cos(kx) + f_2(k) \sin(kx)]$ and then integrate over the surface roughness wavenumber $k$ to obtain the overall behavior. The stability parameter can be calculated in a closed form at a given $k$. The change in the electrochemical potential at a given $k$ is given by: $\Delta \mu_{e^-}(k)=\chi(k) [f_1(k) \cos(kx)+f_2(k) \sin(kx)]$ \cite{ahmad2017-anisotropy}.
This serves to define the stability parameter $\chi(k)$ at a given $k$ as:
\begin{equation}\label{eq:mudefn}
 \chi(k) = \frac{\Delta \mu_{e^-}(k)}{f_1(k)\cos(kx)+f_2(k) \sin(kx)};\\
 \end{equation}
 \begin{equation}\label{eq:mu}
\Delta \mu_{e^-}=-\frac{V_\mathrm{{M}}}{2\mathrm{z}}\left(1 + v \right) \left(-\gamma \kappa 
 - \rtvec{e_n}\cdot [(\rtten{\tau_e} - \rtten{\tau_s})\cdot \rtvec{e_n}]\right)
 + \frac{V_\mathrm{{M}}}{2\mathrm{z}} \left(1 - v \right) \left(\Delta p_e + \Delta p_s \right).
\end{equation}

Eq. \ref{eq:mu} is obtained by including the effect of mechanical stresses and surface tension on the electrochemical potential of the species at a deformed interface \cite{Monroe2004Effect}. $\Delta \mu_{e^-}$ depends on $k$ through the surface tension $\gamma$, curvature $\kappa$, the hydrostatic stress $\Delta p$, and deviatoric stress $\rtten{\tau}$ generated at the interface. $e$ and $s$ in the subscripts refer to the metal anode and solid electrolyte respectively, $V_{\mathrm{M}}$ is the molar volume of the metal atom in the anode, $v$ is the ratio of molar volume of the metal ion in electrolyte $V_{\mathrm{M^{z^+}}}$ to the metal atom in the anode $V_{\mathrm{M}}$, z is the valence of the metal, and $\rtvec{e_n}$ is the normal to the interface pointing towards the electrolyte. The stability parameter consists of contributions from the surface tension and the stresses developed at the metal-electrolyte interface. For an isotropic metal anode with shear modulus $G_e$ and Poisson's ratio $\nu_e$ in contact with an isotropic electrolyte with shear modulus $G_s$ and Poisson's ratio $\nu_s$, the stability parameter $\chi(k)$ can be computed exactly as\cite{ahmad2017stability}:

\begin{eqnarray}\label{eq:chi}
\begin{split}
\chi=&\underbrace{-\frac{\gamma  k^2 V_\mathrm{{M}}(1+v)}{2 \mathrm{z}}}_\text{surface tension} \\
&+ \underbrace{\frac{ 2G_e G_s k V_\mathrm{{M}}(1+v) (\nu_e (4 \nu_s-3)-3 \nu_s+2)}{\mathrm{z} (G_e (\nu_e-1) (4 \nu_s-3)+G_s (4 \nu_e-3) (\nu_s-1))}}_\text{deviatoric stress}\\
&+\underbrace{\frac{k  V_\mathrm{{M}}(1-v) \left(G_e^2 (4 \nu_s-3)+G_s^2 (3-4 \nu_e)\right)}{2 \mathrm{z} (G_e (\nu_e-1) (4 \nu_s-3)+G_s (4 \nu_e-3) (\nu_s-1))}}_\text{hydrostatic stress}
\end{split}
\end{eqnarray}
Using the shear modulus, Poisson's ratio and molar volume ratio of a solid electrolyte, it is possible to calculate the stability parameter for its interface with Li metal anode and determine stability of electrodeposition. For a complete understanding of the interface growth and stability, it is necessary to determine the sign of stability parameter at all the Fourier components $k$. Fortunately, as we will see later, a negative stability parameter at a given $k$ guarantees stability at all higher values.

The molar volume ratio $v=V_{\mathrm{M^{z^+}}}/V_{\mathrm{M}}$ influences the range of shear moduli over which the electrodeposition is stable. $V_{\mathrm{M^{z^+}}}$ was calculated using the coordination number of Li in the crystal structure and mapping them to ionic radius using the values tabulated by Shannon \cite{shannon1976revised}. The coordination number was calculated by generating polyhedra around a species through Voronoi analysis \cite{gotoh1974sphere, Stepanyuk1993micro} as implemented in \texttt{pymatgen} \cite{Ong2013pymatgen}. A linear interpolation was used for computing ionic radius corresponding to coordination numbers not in the Shannon's tabulated values. Predictions with $V_{\mathrm{M^{z^+}}} > V_{\mathrm{M}}$  were ignored since those correspond to very high Li coordination number where Shannon's tabulated values cannot be used. The partial molar volume of the metal in the electrolyte $V_{\mathrm{M^{z^+}}}$ can also be experimentally measured using the potential difference between a stressed and unstressed electrolyte as done by \citealt{pannikkat1999potential} and then using the relationship $V_{\mathrm{M^{z^+}}}=\partial \mu_{\mathrm{M^{z^+}}}/\partial p$ where $\mu_{\mathrm{M^{z^+}}}$ is the electrochemical potential of the metal ion.

Since bulk and shear modulus are related to second derivatives of energy at equilibrium, their calculation by first-principles requires fitting of the energy-strain relationship or the stress-strain relationship. Calculations on several deformed structures are required in order to get an accurate estimate of the fitting parameters. At each deformed state of the structure, the internal coordinates need to be relaxed to calculate the energy or the stress. The \texttt{materials project} employed 24 relax calculations for a single material to compute the moduli. To perform a large scale screening over all Li containing compounds (over 12,000) for use as solid electrolytes, it is necessary to choose a technique that can predict the properties reasonably accurately and without the high computational cost of first-principles simulations.  Hence, we used the crystal graph convolutional neural networks (CGCNN) framework \cite{xie2017crystal} to predict the shear and bulk moduli of the crystalline solid electrolyte materials. At the core of the CGCNN is the multigraph representation of the crystal structure which encodes the atomic information and bonding interactions between atoms. The CGCNN builds a convolutional neural network directly on top of a multigraph that represents the crystal structure of the electrolytes, and predicts the elastic properties by extracting local structural features from the multigraph representation. Note that this method does not depend on any handcrafted geometric or topological features, and all the features are learned by the neural network automatically. This results in a model that is more general than the usual models replying on descriptors but also requires more data to train. 

The training data for the mechanical properties required to compute $\chi$ through Eq. \ref{eq:chi} was obtained from the \texttt{materials project} database \cite{Jain2013, de2015charting}. The calculated values in the database are typically within 15\% of the experimental values which is sometimes the uncertainty in experimental data\cite{de2015charting}.  The moduli have been calculated using density functional theory (DFT) with the Perdew, Becke and Ernzerhof (PBE) Generalized Gradient Approximation (GGA) for the exchange-correlation functional \cite{perdew1996generalized}. GGA-level predictions for 104 systems were within 15\% of the experimental value for all but 16 systems for the bulk modulus and 15 systems for the shear modulus \cite{de2015charting}. Out of the outliers, many had a discrepancy of less than 10 GPa. Experimentally, the shear and bulk moduli can be calculated using the elastic tensor obtained through inelastic neutron scattering or pulse-echo measurements. The experimental measurements typically have a high degree of variability depending on the experimental technique and conditions. We used 2041 crystal structures with shear and bulk moduli, 60\% of the entire dataset with elastic properties, to train our CGCNN model. We performed a data cleaning step in which all materials which were mechanically unstable (116 in number) i.e. with a non-positive definite elastic tensor, were removed from training data. We choose to minimize the mean squared errors between the $\log$ values of predicted and calculated elastic properties since we aim to minimize the relative prediction errors instead of absolute errors and avoid overweighting stiffer materials. This also enabled us to always obtain positive values of the shear and bulk moduli. We then performed a hyperparameter optimization on 20\% validation data via grid search to select the optimum learning rate, weight decay, and number of convolution layers. The best performing hyper-parameters are selected and the resulting model is evaluated on the rest 20\% test data. The CGCNN was implemented in \texttt{PyTorch}~\cite{paszke2017automatic} and the details of the architecture and optimized hyperparameters can be found in the Supporting Information and Ref. \citenum{xie2017crystal}.

\begin{figure}[htbp]
\begin{subfigure}{.45\textwidth}
\includegraphics[scale=0.35]{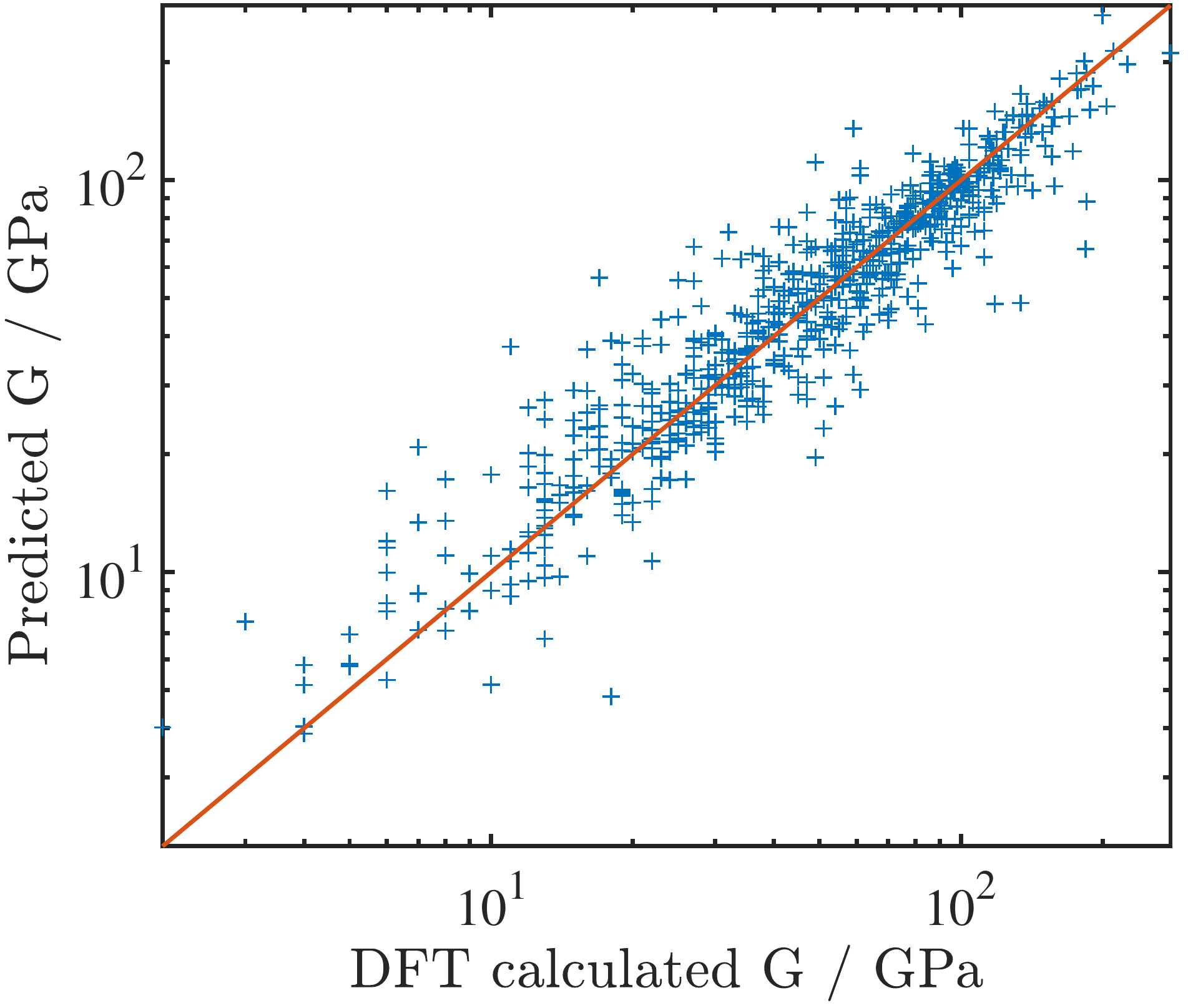}
\caption{\label{fig:pred-shear} Shear modulus $G$.}
\end{subfigure}
\begin{subfigure}{.45\textwidth}
\includegraphics[scale=0.35]{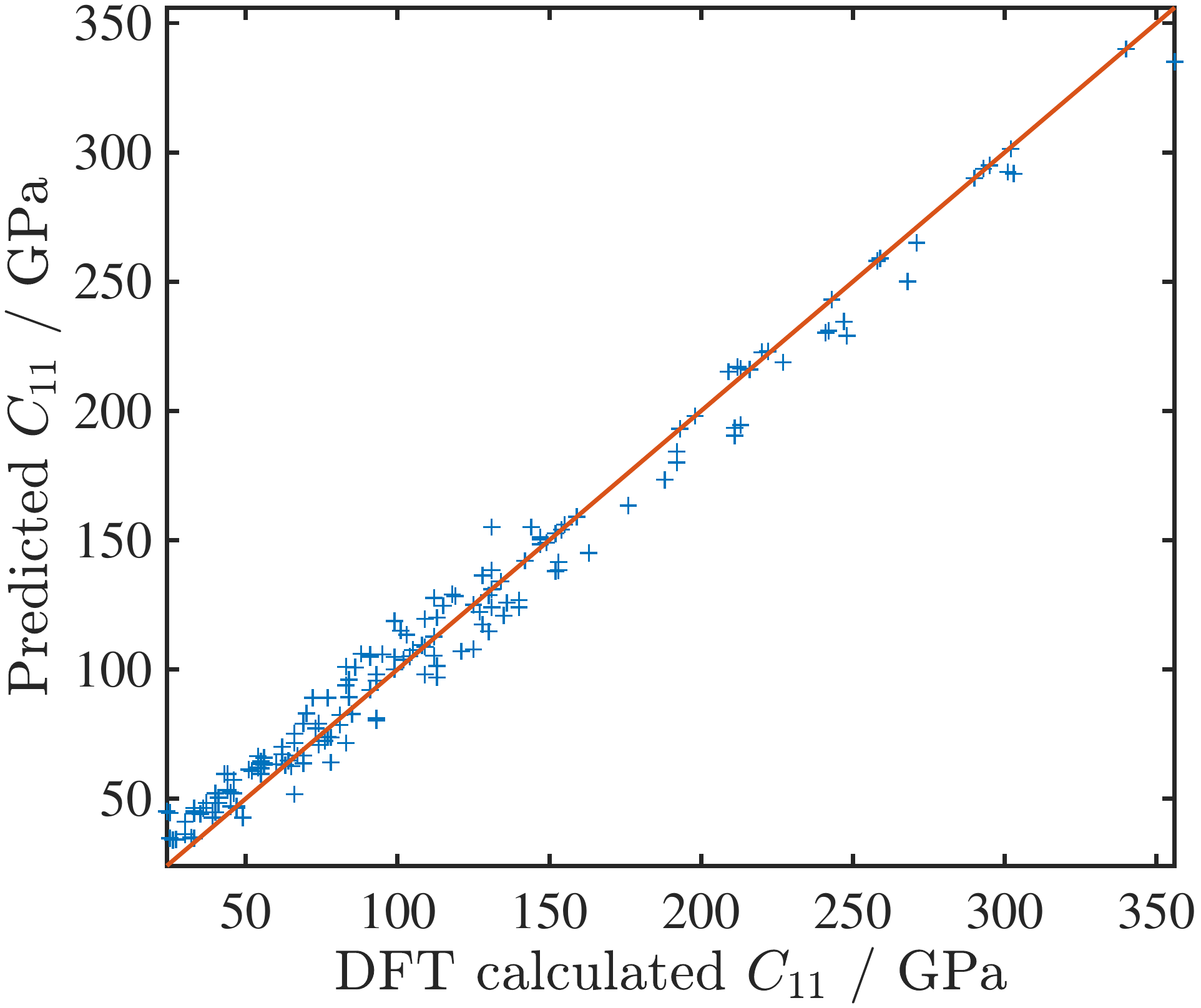}
\caption{\label{fig:pred-c11} Elastic constant $C_{11}$.}
\end{subfigure}
\begin{subfigure}{.45\textwidth}
\includegraphics[scale=0.35]{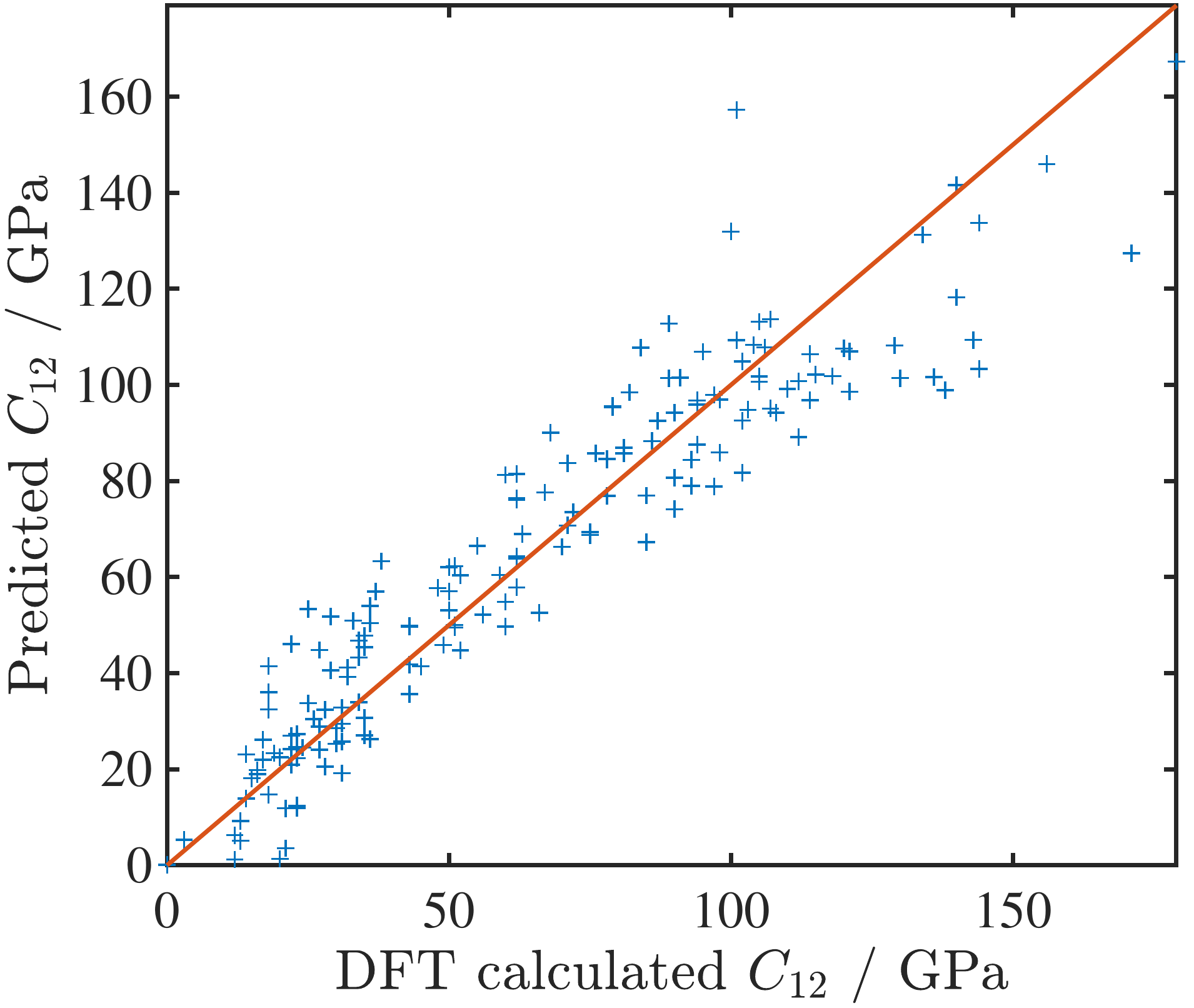}
\caption{\label{fig:pred-c12} Elastic constant $C_{12}$.}
\end{subfigure}
\begin{subfigure}{.45\textwidth}
\includegraphics[scale=0.35]{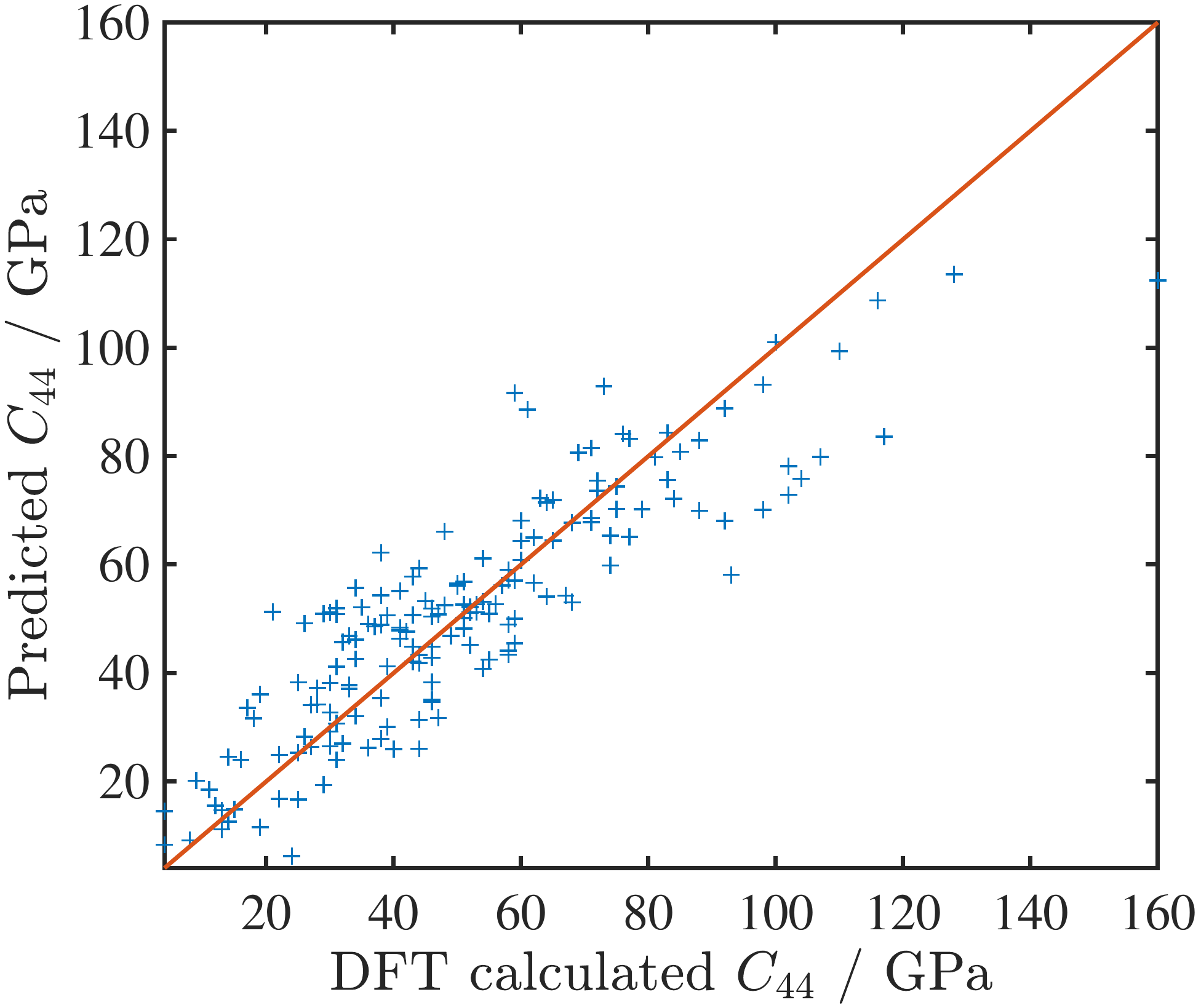}
\caption{\label{fig:pred-c44} Elastic constant $C_{44}$.}
\end{subfigure}
\caption{\label{fig:pred} Parity plots comparing the elastic properties predicted by the machine learning models to the DFT calculated values. The shear modulus is predicted using CGCNN and the elastic constants $C_{11}$, $C_{12}$ and $C_{44}$ are predicted using AdaBoost regression, Lasso regression and Bayesian ridge regression respectively. The parity plot for shear modulus is on 680 test data points while that for the elastic constants contains all available data (177 points). }
\end{figure}

The performance of the CGCNN model was evaluated on 680 test data points. In Figure~\ref{fig:pred-shear}, we show the comparison between the shear modulus predicted by our model and that obtained from the \texttt{materials project} database and in Table \ref{tab:rmse}, we show the RMSE for the shear and bulk moduli predicted by our model.  The RMSE obtained using our model is comparable to previous work by ~\citealt{deJong2016statistical}. However, it is worth noting that we evaluated our model on test data while de Jong et al. evaluated on the entire dataset, indicating that our model does not overfit and has better generalization capability.

\begin{table}[htbp]
\caption{\label{tab:rmse}%
Comparison of RMSE in log(GPa) for shear and bulk moduli}
\begin{tabular}{lcc}
\hline
Method & 
\multicolumn{1}{c}{\textrm{log(G) RMSE}} & 
\multicolumn{1}{c}{\textrm{log(K) RMSE}}\\
\hline
This work & 0.1268 & 0.1013\\
\citealt{deJong2016statistical} & 0.1378 & 0.0750 \\
\hline
\end{tabular}
\end{table}

\begin{figure}[htbp]
\includegraphics[scale=0.5]{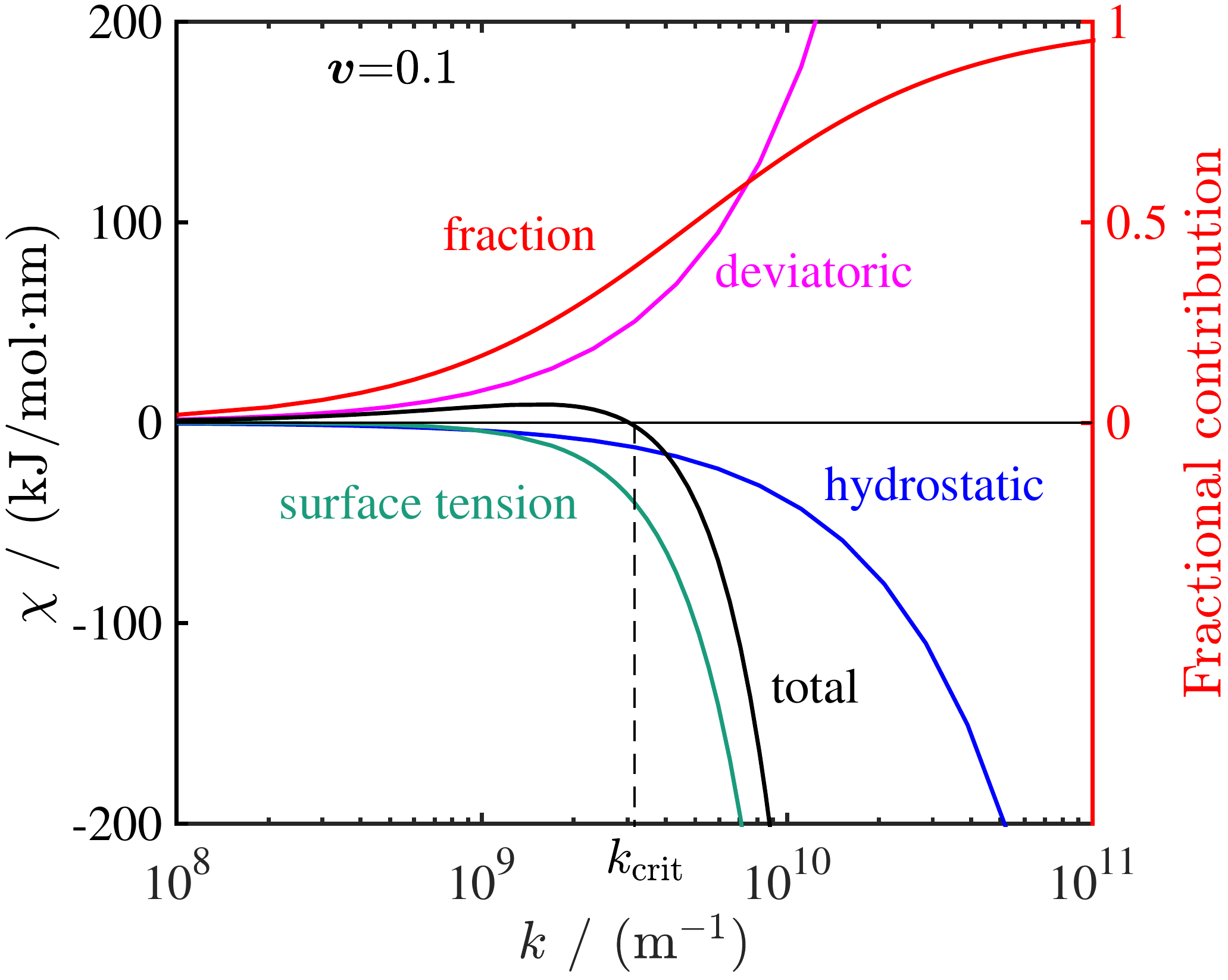}
\caption{\label{fig:chicontrib} Contribution of hydrostatic stress, deviatoric stress and surface tension to the stability parameter as a function of surface roughness wavenumber. The surface tension term starts dominating at high $k$ and ultimately stabilizes the interface after $k=k_{\mathrm{crit}}$. The contributions are plotted for a material with shear modulus ratio $G/G_{\mathrm{Li}}=1$ and Poisson's ratio $\nu=0.33$ which is not stable ($\chi>0$) at $k=10^8$ m$^{-1}$. The red line shows the fraction of surface tension contribution to the stability parameter obtained by dividing the absolute value of its contribution by the sum of absolute values of all components.}
\end{figure}

The shear modulus, Poisson's ratio and molar volume ratio $v=V_{\mathrm{M^{z^+}}}/V_{\mathrm{M}}$ are the parameters determining the stability of electrodeposition at an interface where both materials are isotropic through Eq. \ref{eq:chi}. The role of surface tension in stabilizing electrodeposition is well established \cite{Monroe2004Effect,Tikekar2016,Wang17}. Since the  contribution of the surface tension to the stability parameter increases as $k^2$ while that of stress increases linearly with $k$, the surface tension starts dominating the stability parameter as $k$ is increased. This is elucidated in Fig. \ref{fig:chicontrib} through the contributions of the different terms to the total stability parameter for a material with $G=3.4$ GPa and $v=0.1$. The red line shows the fraction of contribution of surface tension to the overall stability parameter.  All interfaces become stabilized as the value of $k$ is increased beyond the critical surface roughness wavenumber. This motivates a distinction between two types of solid electrolytes -- ones that are stabilized by the stress term alone and those that are stabilized by the surface tension beyond the critical value of $k$. For materials that are stabilized by stresses, the stability parameter remains negative for all values of surface roughness, and therefore, stability is guaranteed. However, for materials that have an overall destabilizing contribution due to stresses (hydrostatic + deviatoric), the stability parameter changes sign at an intermediate value of $k$ since $\chi \to -\infty$ as $k\to \infty$. For such materials, the electrodeposition become stable at the critical surface wavenumber $k_{\mathrm{crit}}=2\pi /\lambda_{\mathrm{crit}}$ (Fig. \ref{fig:chicontrib}). If the surface roughness wavenumber so obtained is possible to achieve by nanostructuring the interface \cite{Wang17}, the electrodeposition might be stabilized.

\begin{figure}[htbp]
\begin{subfigure}{.45\textwidth}
\includegraphics[scale=0.38]{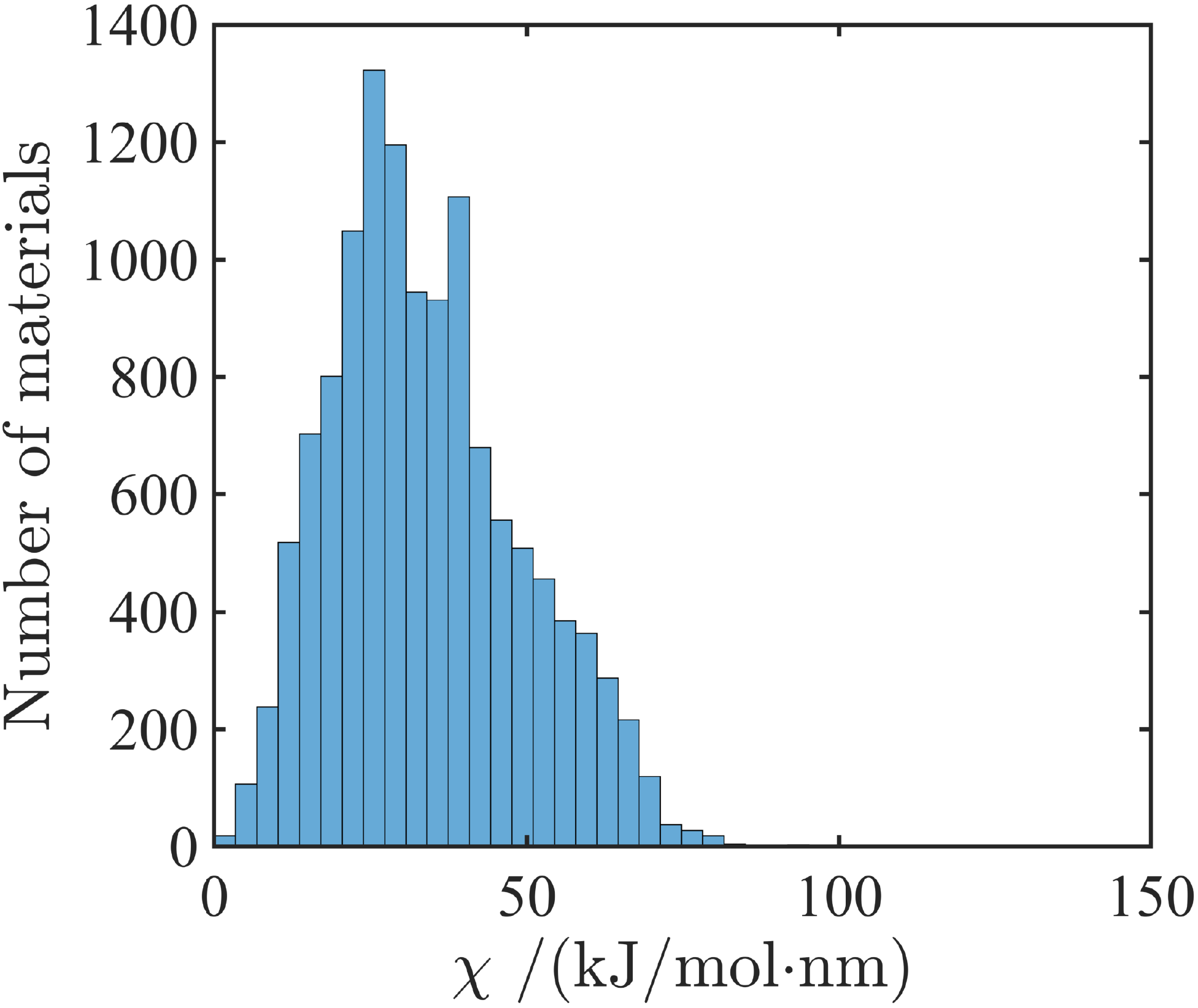}
\caption{\label{fig:chiiso} Distribution of  stability parameter for isotropic Li-solid electrolyte interfaces at $k=10^8$ m$^{-1}$ for 12,950 materials.}
\end{subfigure}
\begin{subfigure}{.45\textwidth}
\includegraphics[scale=0.38]{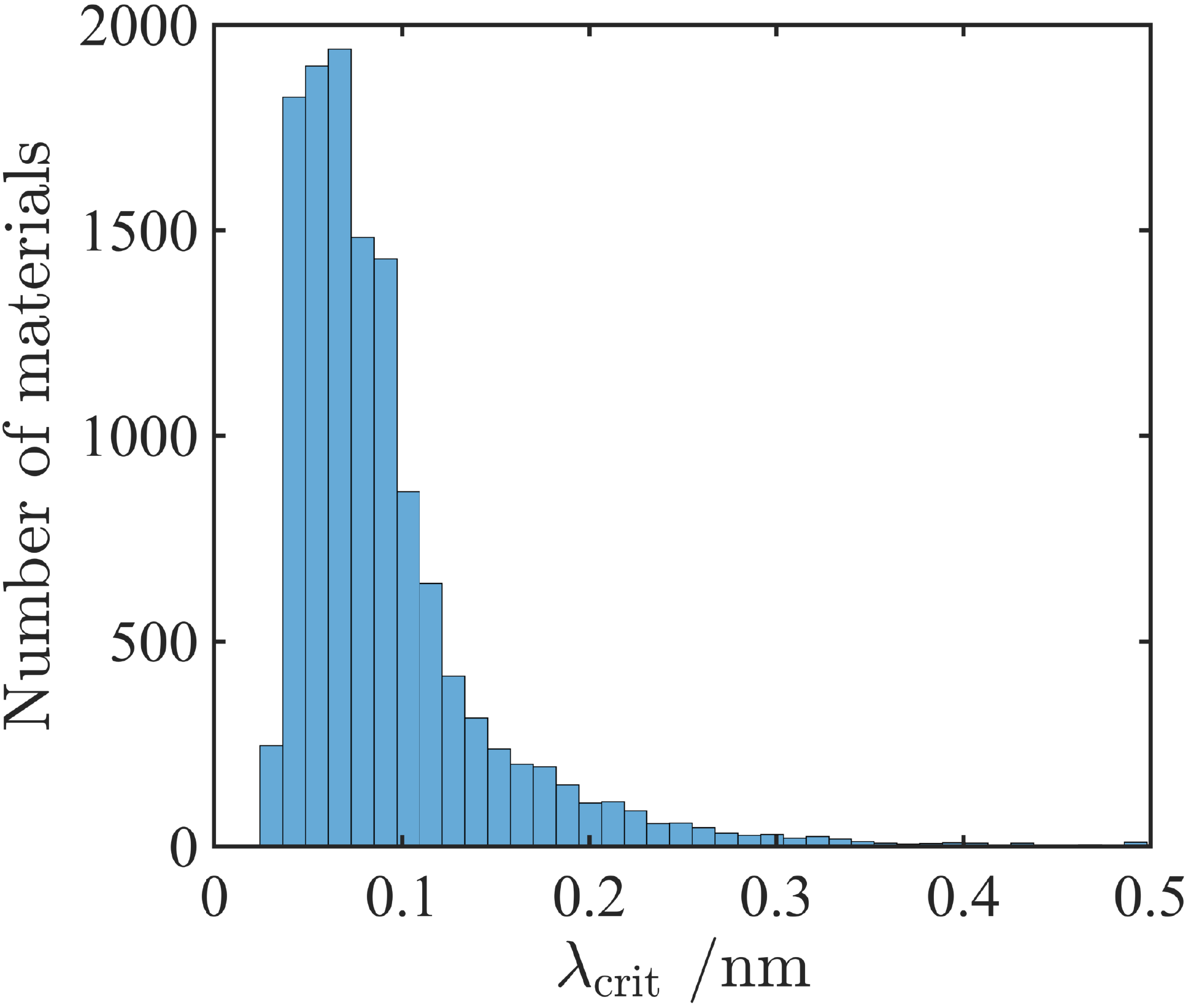}
\caption{\label{fig:lambdaiso} Distribution of critical wavelength of surface roughness for stability for 12,950 materials.}
\end{subfigure}
\caption{\label{fig:chiandlambda} Results of isotropic screening for 12,950 Li containing compounds. None of the materials in the database can be stabilized without the aid of surface tension. The required critical surface roughness wavenumber depends on the contribution of the stress term in the stability parameter.}
\end{figure}

\begin{figure}[htbp]
\includegraphics[scale=0.6]{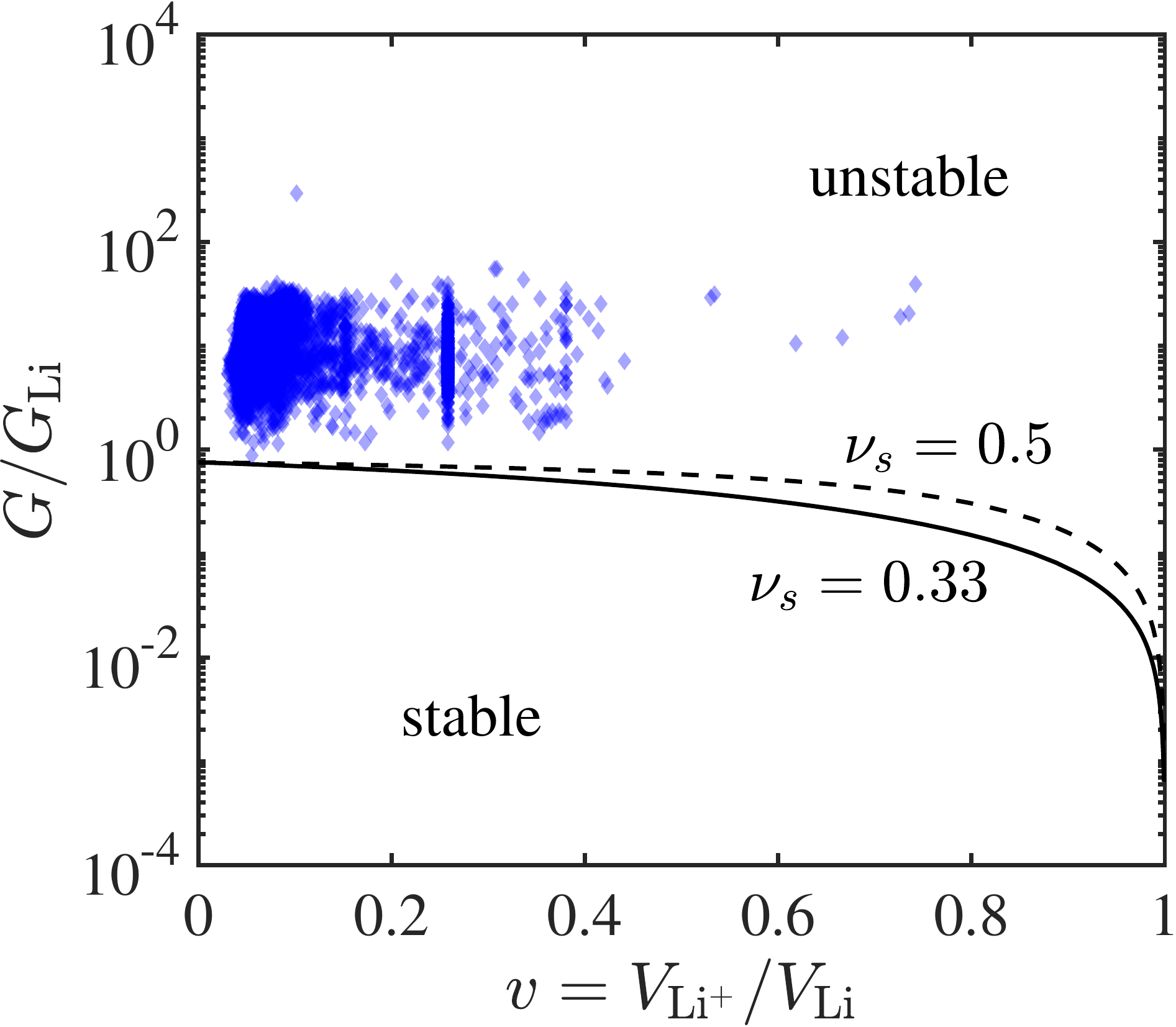}
\caption{\label{fig:stabiso} Isotropic stability diagram showing the position of all solid electrolytes involved in the screening. $G_{\mathrm{Li}}$ is the shear modulus of Li=3.4 GPa. The critical $G/G_{\mathrm{Li}}$ line separating the stable and unstable regions depends weakly on the Poisson's ratio, so the lines corresponding to $\nu_s=0.33$ and 0.5 are good indicators for assessment of stability. The darker regions indicate more number of materials in the region.}
\end{figure}

We computed the stability parameter for 12,950 Li-containing compounds out of which the properties of $\sim$3400 were in training data and those of the remaining were predicted using CGCNN. The stability parameter $\chi$ and the critical surface roughness $\lambda_{\mathrm{crit}}$ for all materials are shown as a histogram in Fig. \ref{fig:chiandlambda}. We found that none of the materials can be stabilized by the stress term alone and all stability parameters are positive at surface roughness wavenumber $k=10^8$ $\mathrm{m}^{-1}$ \cite{Monroe2005Impact}. 
The absence of any materials that can suppress dendrites without assistance from surface tension becomes clear from the isotropic stability diagram shown in Fig. \ref{fig:stabiso}. All materials have $G/G_{\mathrm{Li}}$ ratio higher than the critical value required to stabilize electrodeposition \cite{ahmad2017stability}. The highest number of materials are found in the region where $G/G_{\mathrm{Li}}\sim 15$ and $v\sim 0.1$. The critical wavelength of surface roughening was used as the criteria for screening materials since a higher surface roughness is easier to achieve by nanostructuring. 


The candidate materials with highest critical wavelength of roughening $\lambda_{\mathrm{crit}}$ are shown in Table \ref{tab:iso}. While performing screening, we removed all materials which are electronically conducting i.e. those which have a zero band gap according to \texttt{materials project} database.  However, we retained materials which were thermodynamically metastable since many such solid electrolytes like \ce{Li10GeP2S12} \cite{Kamaya2011} and \ce{Li7P3S11} \cite{yoshikatsu2014LPS} have been successfully synthesized. At nanometer scale roughening, all these materials have negative stability parameter indicating stable electrodeposition. \ce{LiBH4} and \ce{Li2WS4} are 71 meV and 36 meV (P$\bar{4}$2m phase)/31 meV (I$\bar{4}$2m phase) above hull whereas all others in the list are thermodynamically stable. It is worth noting that our screening identifies sulfide and iodide-based solid electrolytes, a class to which many of the current solid electrolytes belong.


\begin{table}[htbp]
\caption{\label{tab:iso} Top interface candidates for stable electrodeposition with Li metal anode together with \texttt{materials project}-id. $\chi$ is the stability parameter which needs to be negative for stability, and $k=2\pi/\lambda$ is the surface roughness wavenumber. Low $k$ corresponds to $k=10^8$ m$^{-1}$ while high $k$ corresponds to a wavelength $\lambda=2 \pi /k=1$ nm. All screened electrolytes can suppress dendrite initiation at high $k$.}
\begin{tabular}{lllccc}
\hline\hline
Formula & Space Group & MP id & \multicolumn{2}{c}{$\chi$/(kJ/mol$\cdot$nm)} & $\lambda_{\mathrm{crit}}/$nm  \\
& & & Low k & High k & \\
\hline
\ce{Li2WS4} & P$\bar{4}$2m & mp-867695 & 0.62 & -109.26 & 3.64\\
\ce{LiAuI4} & P2$_1$/c & mp-29520 & 1.69 & -45.75 & 1.42\\
\ce{Li2WS4} & I$\bar{4}$2m & mp-753195 & 1.75 & -38.54 & 1.34\\
\ce{LiBH4} & P$\bar{1}$ & mp-675926 & 1.98 & -40.13 & 1.32\\
\ce{LiGaI4} & P2$_1/$c & mp-567967  & 1.81 & -35.99 & 1.31\\
\hline
\end{tabular}
\end{table}


\subsection{Anisotropic Material Screening}
The isotropic assumption provides a good starting point for determining the stability of electrodeposition in terms of the mechanical and chemical properties of the solid electrolyte. However, this assumption does not take into account the huge anisotropy of Li metal resulting in an anisotropy index of 8.52\cite{xuLi2017}. This may also arise when the surface of the the solids in contact are single crystalline or dominated by a particular crystallographic orientation as has been seen for Li metal\cite{Cui2017-texturing}. Therefore, a full anisotropic treatment of the mechanics taking into account the elastic tensor and orientation of Li metal at the interface is essential. Previously, we also observed huge variation in the stability parameter, with the \hkl[010] orientation at the interface being the most compliant \cite{ahmad2017-anisotropy}. 
Here, we analyze the stability of electrodeposition at an anisotropic interface involving crystalline solid electrolyte in contact with crystalline Li metal.

For this analysis, we calculate the stability parameter using the full elastic tensor of the electrode and the electrolyte. A knowledge of surface orientations at the interface is necessary to determine the elastic tensor. Specifically, we considered \hkl<1 0 0>, \hkl<1 1 0>, \hkl<1 1 1> and \hkl<2 1 1> crystallographic directions of Li normal to interface, in contact with seven low index facets of the solid electrolyte for the screening procedure. The last three crystallographic directions have been observed through cryo-electron microscopy as the dominant directions for Li dendrite growth in carbonate-based electrolytes  \cite{Cui2017-cryo}. Additionally, X-ray diffraction pole figure analysis has revealed that Li gets electrodeposited along certain preferential crystallographic orientations \cite{Cui2017-texturing}. The crystallographic orientation provides one more parameter for tuning the mechanical properties at the interface. In an earlier work, we used the Stroh formalism \cite{stroh1958,stroh1962} to calculate the stresses and deformations at the interface which were used to obtain the stability parameter $\chi$\cite{ahmad2017-anisotropy}.

We developed a model different from CGCNN for predicting the full elastic tensor of each material since the size of training data is much smaller for the full elastic tensor. Considering the success of atomistic descriptors such as Li neighborhood, anionic framework \cite{wang2015design}, bond ionicity etc. in predicting Li ion conduction \cite{sendek2017holistic}, we developed a regression model for predicting the elastic tensor of all Li-containing compounds using several of these relevant descriptors.  A complete list of descriptors can be found in the Table S2 of Supporting Information. All descriptors used in the model are geometric or topological and can be obtained from the structure of the material in CIF format \cite{Hall1991CIF}. We employed the crystal symmetries to reduce the number of parameters in the elastic tensor. This procedure reduces the number of constants in the elastic tensor to train from 21 to 3 in the cubic crystal class for example and by construction, ensures the predicted elastic tensor obeys its crystal symmetry\cite{coudert2014elastic}. We trained separate models for each elastic constant, which were used to build the full elastic tensor of the material. We believe the differences in the uncertainty of the different elastic constants\cite{Ahmad16uncertainty} results in different amounts of noise in the training data, thereby, justifying the need for separate models.

The training data for the elastic tensor was obtained from \texttt{materials project} database \cite{Jain2013, de2015charting} retrieved using \texttt{pymatgen} \cite{Ong2013pymatgen}, with the same level of theory as the isotropic case. Since sufficient data was available only for the cubic crystal class (see Table S3 of Supporting Information), we developed a regression model for predicting the elastic tensor of only cubic Li-containing compounds. However, our framework is very general and can accommodate more data as it is generated in training set as well as add models for other symmetries once the size of training data is sufficient in \texttt{materials project} database. For other crystal symmetries besides cubic, we screened over materials with elastic tensor available from first-principles. Our screening process consisting of DFT-calculated data for all symmetries and predicted data for cubic symmetries covers >15,000 Li metal-solid electrolyte interfaces. 


The regression model was chosen on the basis of cross validation performance (coefficient of determination) on the values of the elastic constants (in GPa). The models tested for performance were Lasso regression, elastic net, kernel ridge regression, Gaussian process regression, gradient boosting regression (GBR) \cite{friedman2001gbr,friedman2001elements} used in an earlier work on predicting mechanical properties of zeolite frameworks \cite{evans2017predict}, AdaBoost regression \cite{freund1997boost,drucker1997boost}, support vector regression (SVR) \cite{}, random forest regression, Bayesian ridge regression \cite{mackay1992bayesian}, least angle regression (LARS), and automatic relevance determination (ARD). All hyperparameters were optimized using grid search with nested 3-fold cross validation. The optimized hyperparameter values can be found in the Supporting Information. The \texttt{scikit-learn} package \cite{pedregosa2011scikit}, which has implementations of all models listed above, was used for training, hyperparameter optimization and prediction.

Figures \ref{fig:pred-c11}, \ref{fig:pred-c12} and \ref{fig:pred-c44} show the comparison of predicted and calculated values of elastic constants $C_{11}$, $C_{12}$ and $C_{44}$. The best models giving the least cross validation error were: AdaBoost regression for $C_{11}$, Lasso regression for $C_{12}$ and Bayesian ridge regression for $C_{44}$. We obtain an overall $R^2$ of 0.98, 0.85 and 0.69 and a cross-validation $R^2$ of 0.68, 0.85 and 0.71 respectively using 3-fold cross validation. The dominant atomistic features that determined the elastic tensor were: volume per atom, packing fraction and anion-anion separation distance. As expected, the material stiffness decreases with volume per atom. The stiffness increases with increase in minimum anion-anion separation showing the dominant role of anion-anion interactions in the crystal. This indicates a flatter energy landscape at lower anion-anion separation distance. The stiffness increases with the packing fraction of sublattice consisting of all atoms in the structure but Li and decreases with increase in overall packing fraction of the crystal.

While evaluating the models, the inherent uncertainty in DFT calculated values due to exchange correlation functional and error with respect to experimental data should be taken into account~\cite{Ahmad16uncertainty,Deng01012016}. The uncertainty and the errors in DFT calculated values generate noise in the training data. For example, DFT calculated $C_{12}$ generally has high uncertainty~\cite{Ahmad16uncertainty} which appears to have some signature in its cross-validation performance as well as the simple model that provides the least cross validation error for it: Lasso. Similarly, $C_{44}$ has low uncertainty but high errors with respect to experimental data\cite{Deng01012016} which generates noise in its training set. The best model for $C_{44}$ is Bayesian ridge regression, a linear model. $C_{11}$ has comparatively lower uncertainty and error which makes it's training data better suited to the use of more sophisticated models. The best model for $C_{11}$ is AdaBoost regression whose performance is well known to degrade in the presence of noisy data and outliers due to the use of exponential loss function\cite{friedman2001elements}. We believe the lack of noise and outliers in the data for $C_{11}$ makes AdaBoost a better candidate for training and prediction than linear models. Overall, our models have satisfactory performance to be used for the next step of prediction and screening. 


Our anisotropic screening procedure involves the determination of electrodeposition stability for 482 electrolyte  materials with DFT-computed elastic tensor from \texttt{materials project} database and 548 materials with cubic crystal structure whose elastic tensor was predicted using the regression model developed earlier. The total number of Li-solid electrolyte interfaces involved in the screening is over 15,000 (see also Table S3 in Supporting Information). The best candidate interfaces obtained from the screening procedure are shown in Table \ref{tab:aniso}. It is worth pointing out that all the screened candidates are stabilized by the stress term in the stability parameter and don't depend on surface tension for stability. We observe two major features in the materials obtained through screening: either the material is mechanically soft i.e. with small eigen values of the elastic tensor or the material has highly anisotropic mechanical properties. The mechanical softness of these materials is particularly attractive since it means the mechanical properties for dendrite suppression and ionic conductivity can be optimized simultaneously through methods like increasing volume per atom\cite{wang2015design}. The high anisotropy results in certain weak directions along which the electrolyte may comply with the Li metal anode. This degree of freedom in crystallographic orientation is probably the reason why there are several screened candidates in anisotropic case compared to the isotropic case. All screened candidates in Table \ref{tab:aniso} have a universal anisotropy index greater than 10, which is zero for an isotropic material \cite{ranganathan2008anisotropy}. The highest anisotropy index of 100.84 and 66.07 is found for \ce{LiNdHg2} and \ce{LiDyTl2} while the two phases of \ce{Li2WS4} have the values 31.30 and 12.84.   The mechanically soft nature of these materials is reminiscent of density-driven stability in the isotropic case for materials with $v<1$. \cite{ahmad2017stability}.


\begin{table}[htbp]
\caption{\label{tab:aniso} Top twenty interfaces for stable electrodeposition with Li metal anode with their \texttt{materials project} id, interface normal and stability parameter. The last column shows the universal anisotropy index $A^U$ which is zero for a completely isotropic material~\cite{ranganathan2008anisotropy}.}
\begin{tabular}{lllcccr}
\hline\hline
Formula & Space Group & MP id & \multicolumn{2}{c}{Interface normal} & $\chi$@$k=10^8$ $\mathrm{m^{-1}}$ & $A^U$ \\
& & & Li & Electrolyte & (kJ/mol$\cdot$nm)\\
\hline
\ce{Li2WS4} & P$\bar{4}$2m & mp-867695 & \hkl[1 1 1] & \hkl[0 0 1] &  -1.92 & 31.30\\
\ce{Li2WS4} & P$\bar{4}$2m & mp-867695 & \hkl[2 1 1] & \hkl[0 0 1] &  -1.87 & 31.30\\
\ce{Li2WS4} & P$\bar{4}$2m & mp-867695 & \hkl[0 1 0] & \hkl[0 0 1] &  -1.68 & 31.30\\
\ce{Li2WS4} & I$\bar{4}$2m & mp-753195 & \hkl[0 1 0] & \hkl[0 0 1] &  -1.23 & 12.84\\
\ce{LiBH4}  & P$\bar{1}$ & mp-675926 & \hkl[0 1 0] & \hkl[0 1 0] & -1.12 & 13.65\\
\ce{Li2WS4} & I$\bar{4}$2m  & mp-753195 & \hkl[1 1 1] & \hkl[1 0 1] & -1.00 & 12.84\\
\ce{LiOH} & P4/nmm & mp-23856 & \hkl[0 1 0] & \hkl[0 0 1] &  -1.00 & 113.29\\
\ce{Li2WS4} & I$\bar{4}$2m   & mp-753195 & \hkl[1 1 1] & \hkl[0 0 1] &   -1.00 & 12.84\\
 \ce{LiOH} & P4/nmm & mp-23856 & \hkl[2 1 1] & \hkl[0 0 1] & -0.99 & 113.29\\
 \ce{LiOH} & P4/nmm & mp-23856 & \hkl[1 1 1] & \hkl[0 0 1] & -0.98 & 113.29\\
\ce{Li2WS4} & I$\bar{4}$2m  & mp-753195 & \hkl[2 1 1] & \hkl[0 0 1] & -0.89 & 12.84\\
\ce{Li2WS4} & I$\bar{4}$2m  & mp-753195 & \hkl[0 1 0] & \hkl[1 0 1] & -0.79 & 12.84\\
\ce{LiBH4} & P$\bar{1}$  & mp-675926 & \hkl[1 1 1] & \hkl[1 1 0] & -0.77 & 13.65\\
\ce{LiBH4} & P$\bar{1}$ & mp-675926 & \hkl[1 1 1] & \hkl[0 1 0] &  -0.75  & 13.65\\
\ce{Li2WS4} & I$\bar{4}$2m & mp-753195 & \hkl[0 1 0] & \hkl[0 1 1] &  -0.49 & 13.84\\
\ce{LiBH4} & P$\bar{1}$ & mp-675926 & \hkl[0 1 0] & \hkl[1 1 0] &  -0.47 & 13.65\\
\ce{LiNdHg2} & Fm$\bar{3}$m & mp-861660 & \hkl[1 1 1] & \hkl[111] &   -0.46 & 110.84\\
\ce{LiDyTl2} & Fm$\bar{3}$m & mp-867108 & \hkl[1 1 1] & \hkl[111] &   -0.41 & 66.07\\
\ce{Li2WS4} & P$\bar{4}$2m & mp-867695 & \hkl[1 1 0] & \hkl[001] &   -0.40 & 31.30\\
\ce{LiNdHg2} & Fm$\bar{3}$m & mp-861660 & \hkl[2 1 1] & \hkl[1 1 1] &   -0.29 & 110.84\\
\hline
\end{tabular}
\end{table}

Although literature on the candidates we obtained through screening is rather limited, we do find some similarities with solid electrolyte materials currently being explored. For example, Li borohydrides have been recently explored as fast ion conductors for solid state batteries \cite{ziheng2017borohydrides,varley2017polyborane} Their anion substituted analogues also exhibit high conductivities and provide opportunities for tuning other desirable properties  such as electrochemical and thermodynamic stability\cite{tang2015liquid,varley2017polyborane, tang2015unparalleled}. \ce{Li2WS4} phases are sulfides with structure similar to the tetragonal phase of \ce{LiFePO4}. One of the candidates, \ce{LiOH}, is often found as an electrochemically stable product in the solid electrolyte interphase formed at the anode\cite{Malmgren2013SEI,tasaki2009solubility}. We expect \ce{LiNdHg2} to be likely toxic while \ce{LiDyTl2} is expected to be expensive due to low abundance of constitutive elements.  The candidates we obtained through screening will only be useful if they satisfy several other requirements imposed by solid electrolytes.  Here we discuss the other properties of the screened candidates for enabling their use as solid electrolytes:

\emph{Thermodynamic stability}: Besides \ce{LiBH4} and \ce{Li2WS4}, all screened candidates are thermodynamically stable i.e. energy above the convex hull is 0 eV. The two phases of \ce{Li2WS4} are 31 and 36 meV above hull. The hexagonal phase of \ce{LiBH4} has been explored previously as a solid electrolyte with promising performance~\cite{motoaki2007libh4,maekawa2009libh4, Das2016LiBH4}. This phase is thermodynamically unstable at room temperature. However, doping with Li halides improved the stability of the hexagonal phase even below room temperature \cite{maekawa2009libh4}. It is possible that a similar scheme might help stabilize the screened triclinic P$\bar{1}$ phase obtained through our screening as well.  

\emph{Electronic conductivity}: All the screened solid electrolytes in the list are electronic insulators since we removed those with a zero GGA band gap. All candidates have band gap > 1 eV except \ce{LiNdHg2} and \ce{LiDyTl2} which have gaps of 14 meV and 172 meV. Since band gaps are generally underestimated by GGA level DFT, we expect the actual gaps to be much higher.

\emph{Ionic conductivity}: In general, we expect the high anisotropy of mechanical properties to affect the activation energy landscape for Li ion conduction as well, possibly leading to anisotropic conductivities. The high temperature hexagonal phase of \ce{LiBH4} has a Li ion conductivity as high as 10 mS cm$^{-1}$\cite{motoaki2007libh4}. The screened triclinic P$\bar{1}$ phase of \ce{LiBH4} is disordered and is expected to have high Li ionic conductivity. It will be worth investigating the possibility of tuning its ionic conduction and mechanical properties through anion substitution similar to its hexagonal analogue \cite{tang2015unparalleled,tang2015liquid}. The two phases of \ce{Li2WS4} have a high volume per sulfur atom which may aid Li ion conduction \cite{wang2015design}. \ce{LiOH} is known to undergo a solid state transition at $\sim$415 $^\circ$C accompanied by a large increase in ionic conductivity\cite{johnson1977ionic}.

We found several candidates that could enable dendrite suppression through the anisotropic criteria.  It is worth highlighting that grain misorientation needs to be avoided for these materials.  Textured growth of Li film on Li metal anode has been observed for liquid electrolytes \cite{Cui2017-texturing} and so this might present a real opportunity.  A similar opportunity for textured growth of Li metal is present with solid electrolytes, however, it would require precise engineering of film growth at the interphase which requires further investigation.

While the candidate list identified is small, we strongly believe that there is a lot of room for further search for candidates.  In particular, we find that anisotropy plays a crucial role in determining the stability and thus, there may be additional non-cubic Li-containing compounds that could suppress dendrites at certain crystallographic orientations.  Further, the degree of disorder is another important factor and many glassy or amorphous materials are known to be good solid electrolytes.  These avenues will be the subject of our future investigations.

\section{Conclusions}
We applied criteria for stable electrodeposition together with machine learning techniques to computationally screen solid electrolytes for suppressing Li dendrite growth. The machine learning techniques accelerate the process of screening by predicting the properties of solid electrolytes through the identification of structure-property relationships. We train a convolutional neural network on shear and bulk moduli. We employ AdaBoost, Lasso and Bayesian ridge regression for training the elastic constants $C_{11}$, $C_{12}$ and $C_{44}$, respectively, of cubic materials. The choice of machine learning models used was rationalized by the model's ability to handle noise in the training data. Our approach is readily applicable for screening materials for other properties of interest and can easily accommodate more data as it is generated.

Our predictive models enabled us to screen 12,950 solids using isotropic stability criteria and over 15,000 interfaces using anisotropic stability criteria of electrodeposition on the Li metal anode. In the isotropic case, we found no materials that could be stabilized solely by the stresses generated at the interface; however, a surface tension-mediated stabilization was found to be possible at high surface roughness wavenumbers. In the anisotropic case, the additional degree of freedom related to crystallographic orientation of the solid electrolyte at the interface enabled us to find over twenty interfaces with six solid electrolytes that are predicted to be stable to dendrite initiation. We identify some common features like anisotropy and mechanical softness present in the screened candidates based on which one can simultaneously optimize properties required for dendrite suppression as well as fast ion conduction. We believe that the use of techniques like doping and defect generation will be crucial to ensure simultaneous satisfaction of other solid electrolyte requirements for potential candidates. Experimental studies on the properties and performance of these candidate solid electrolytes can provide major insights into the dendrite suppression mechanism. 

\begin{acknowledgement}

ZA thanks A. Sendek for sharing data on features of Li-containing compounds, and S. Shekhar, C. Hwu, L. Kara, J. Montoya, K. Thomas-Alyea for helpful discussions. Z.A. acknowledges support from the Advanced Research Projects Agency-Energy (ARPA-E) Integration and Optimization of Novel Ion Conducting Solids (IONICS) program under Grant No. DE-AR0000774. V.V. gratefully acknowledges support from the U.S. Department of Energy, Energy Efficiency and Renewable Energy Vehicle Technologies Office under Award No. DE-EE0007810. This work used the Extreme Science and Engineering Discovery Environment (XSEDE)\cite{towns2014xsede}, which is supported by National Science Foundation Grant No. OCI-1053575.  Specifically, it used Grant No. MSS170010P on the Bridges system\cite{nystrom2015psc}, which is supported by NSF Award No. ACI-1445606, at the Pittsburgh Supercomputing Center (PSC).

\end{acknowledgement}

\begin{suppinfo}
The following files are available free of charge.
\begin{itemize}
\item suppinfo.pdf: Details of machine learning models
  \item  shear-modulus.xlsx: shear modulus of 60,648 compounds predicted using CGCNN
  \item bulk-modulus.xlsx: bulk modulus of 60,648 compounds predicted using CGCNN
 \item train-chi.xlsx: anisotropic stability parameter of 482 Li-containing compounds with property data from \texttt{materials project} database.
  \item pred-chi.xlsx: anisotropic stability parameter of 548 cubic Li-containing compounds with properties predicted using our model
\end{itemize}

\end{suppinfo}

\bibliography{refs}

\includepdf[pages=1-7]{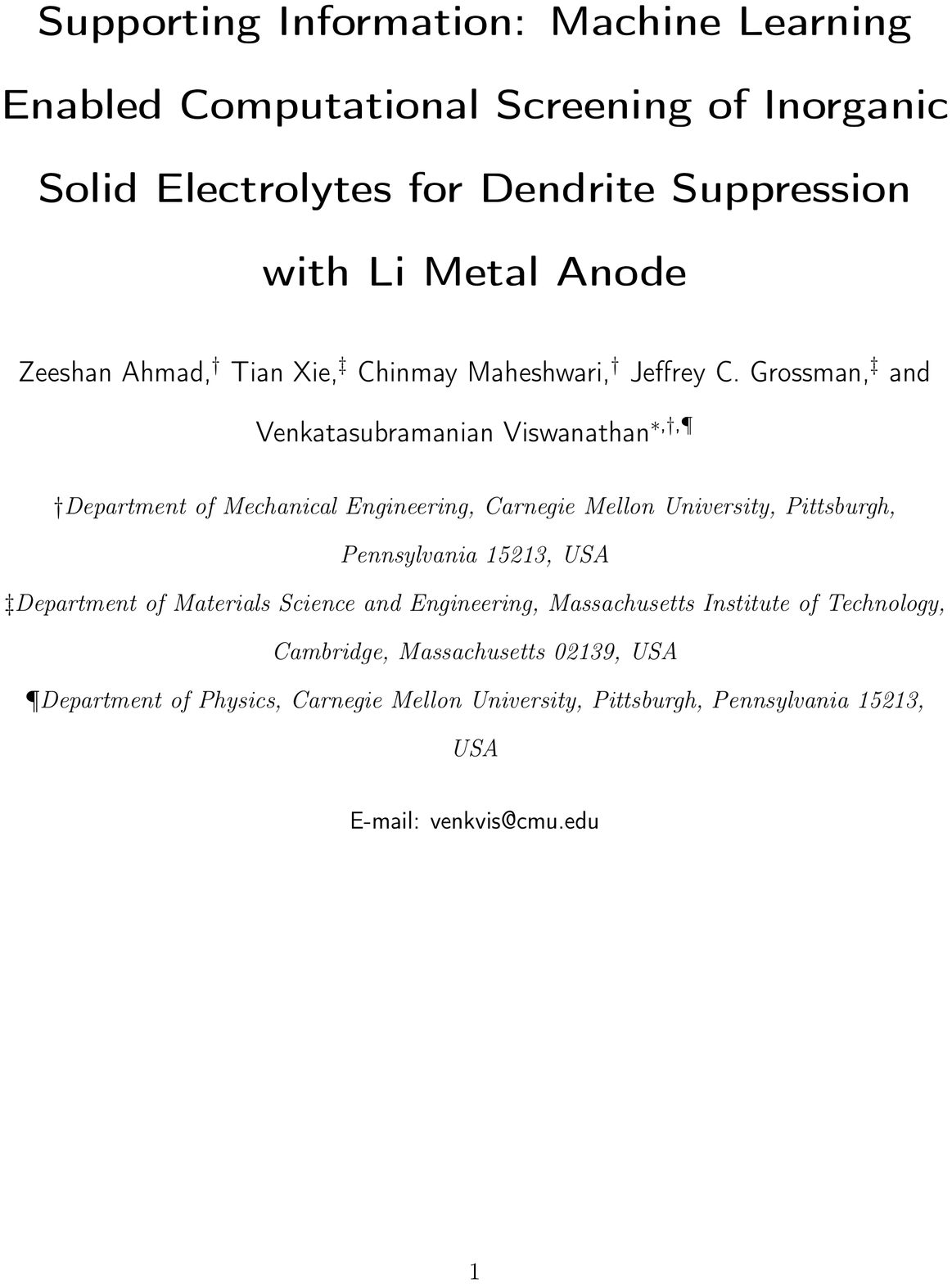}
\end{document}